\documentstyle[twocolumn,seceq,epsf]{jpsj}

\def\runtitle{Quantum Dynamics in Nanoscale Magnets in 
Dissipative Environments}
\def\runauthor{ Seiji {\sc Miyashita}, et al.}

\title
{ Quantum Dynamics in Nanoscale Magnets in 
Dissipative Environments}

\author
{ Seiji {\sc Miyashita}\footnote{E-mail address: 
miya@yuragi.t.u-tokyo.ac.jp},
Keiji{\sc Saito}\footnote{E-mail address: 
saitoh@spin.t.u-tokyo.ac.jp}
Hiroto{\sc Kobayashi}\footnote{E-mail address:
hiroto@spin.t.u-tokyo.ac.jp}
and Hans {\sc De Raedt}\footnote{E-mail address:
H.A.de.Raedt@phys.rug.nl}
}

\inst
{
Department of Applied Physics, The University of Tokyo, 
Hongo 7-3-1, Bunkyo-ku, Tokyo 113-8656, Japan
}

\recdate
{
\today
}

\abst
{In discrete energy structure of nanoscale magnets, 
nonadiabatic transitions at avoided level crossings lead to 
fundamental processes of dynamics of magnetizations. 
The thermal environment causes dissipative 
effects on these processes. 
In this paper we review the features of the nonadiabatic transition 
and the influence of the thermal environment. 
In particular we discuss the temperature independent stepwise structure 
of magnetization at very low temperatures 
(deceptive nonadiabatic transition), the alternate enhancement 
of relaxation in the sequence of resonant tunneling points 
(parity effects), and processes caused by combinations 
of nonadiabatic transitions and disturbance due to external noises.
}

\kword
{Nanoscale magnets,nonadiabatic transitions,
Landau-Zener-St\"uckelberg transitions,dissipative environments}

\begin{document}
\sloppy
\maketitle

\section{Introduction}

Hysteresis phenomena of ferromagnets have been 
one of the most interesting problems in the magnetism and statistical physics. 
Mechanism of the coercive force has been investigated by 
studying the processes that lead to the critical nucleation 
and motion of the domain wall.\cite{L6769,Fisher} 
From the point of view of free energy of the system, 
the hysteresis phenomena have been discussed 
in terms of the relaxation process of 
the metastable state to the true equilibrium state 
in the picture shown in Fig. 1 (a).
Usually the end point of hysteresis is related to 
the so-called spinodal point 
where the metastability disappears (see Fig. 1 (b)).

\begin{figure}
\noindent
\epsfxsize=7.8cm \epsfysize=5.0cm \epsfbox{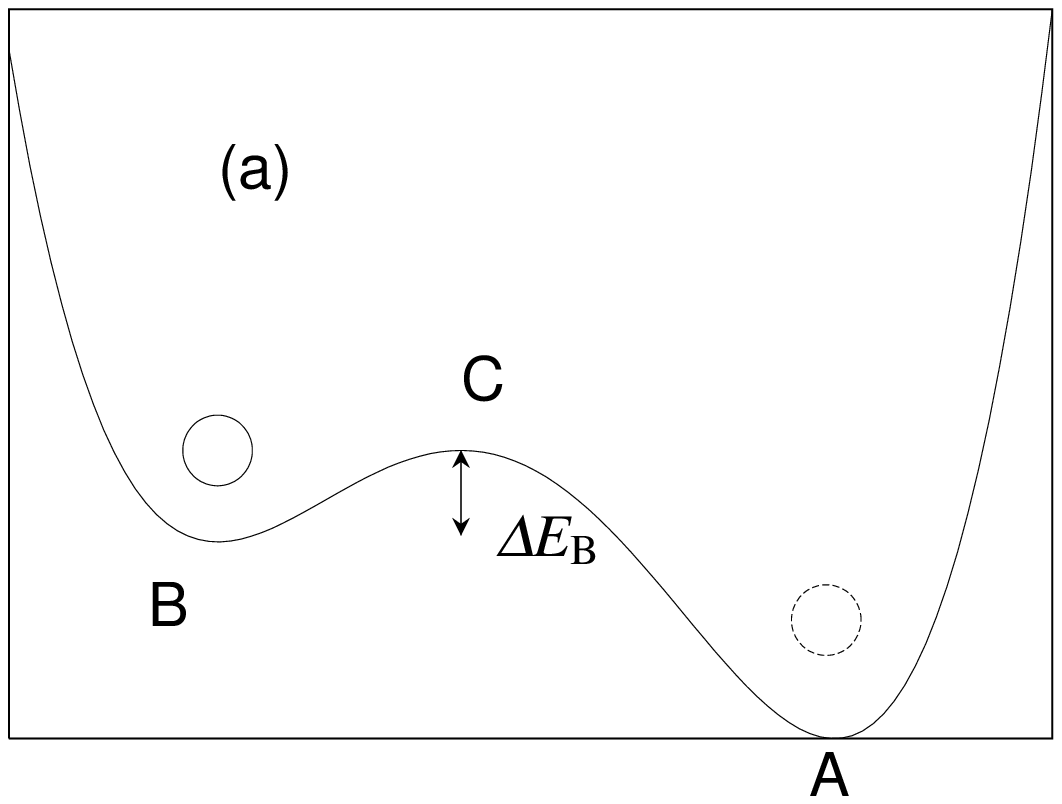} \\
\epsfxsize=7.8cm \epsfysize=5.0cm \epsfbox{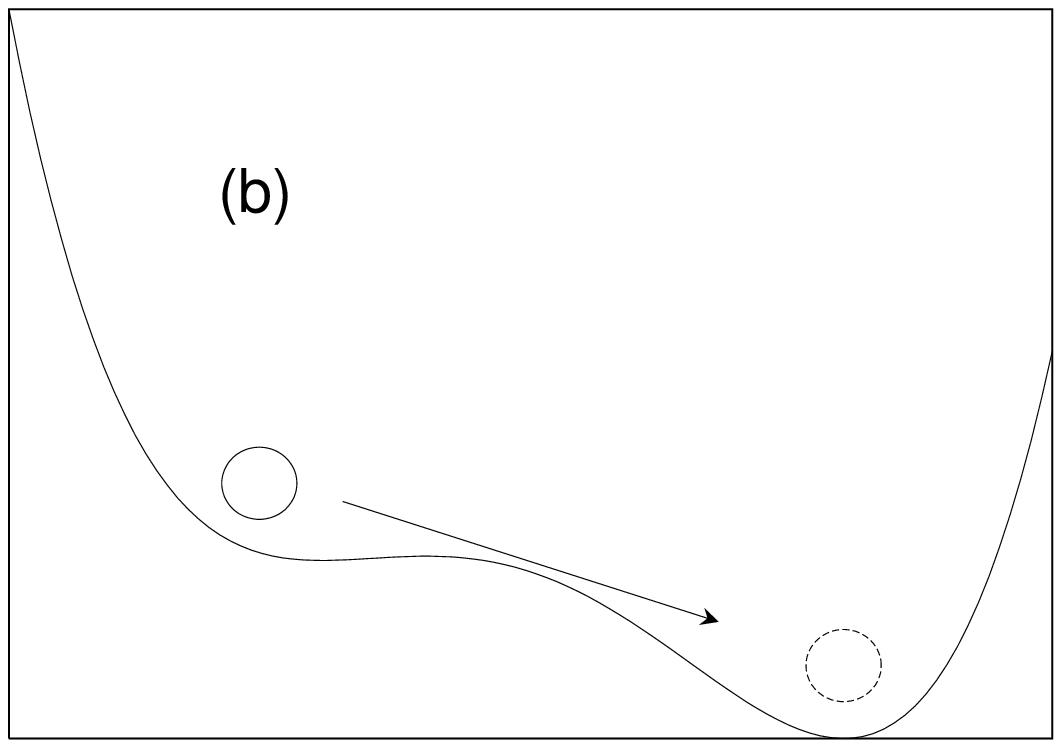}
\label{fig1}
\caption{Potential pictures of relaxation: (a) metastable (b) spinodal.}
\end{figure}

However there is some probability of relaxation 
from the metastable state B to the equilibrium state A.
For example at finite temperatures the probability 
of thermal excitation to the top of barrier C 
is proportional to $\exp(-\Delta E_{\rm B}/k_{\rm B}T)$ 
and thus the relaxation rate  
$p_{\rm th}$ of the metastable state through this activation process
is given by 
\begin{equation}
p_{\rm th}={1\over\tau_0}e^{-\Delta E_{\rm B}\over k_{\rm B}T} ,
\label{p-TH}
\end{equation}
i.e., the Arrhenius law. 
At low temperatures the relaxation time 
$\tau=1/p_{\rm th}$ of this process diverges exponentially. 

The shapes of the free energy in Fig. 1 are given by the mean field theory, 
which gives a good intuitive picture of the metastability. 
Here it should be noted that 
$\Delta E_{\rm B}$ should be a microscopic quantity. 
Because in a bulk system
$\Delta E_{\rm B}$ is of order the system size,
the activation rate vanishes, i.e., $p_{\rm th} = 0$. 
For a more quantitative understanding, 
we must look at the system microscopically.
There the system is not uniform and 
we have to consider 
a microscopic break through 
of the metastable state.
Such break through occurs as a process of 
creating the critical nucleus.\cite{Fisher}
For this microscopic process,
Figs. 1(a)-(b) represent effective potentials of the size of 
the nucleus.
The relaxation time of metastable states has been classified 
according to the size of system 
and the generating rate of the critical nuclei. 
There are two regions, i.e.,
a single nucleation region (stochastic region) 
and a multi nucleation region (Avrami region).\cite{RTMS94}

When the size of the magnets becomes smaller 
than the width of the domain wall, 
the nucleus can not be defined. 
In such cases, the magnetization of the system changes
uniformly and this process of breakdown of the metastable state
is called "coherent process". 
Relaxations in this situation also 
have been studied extensively.\cite{coherent}

It has been pointed out that quantum fluctuation may play an 
important role in such small systems.
In order to detect such quantum processes, 
several experiments have been proposed.\cite{ferritin} 
However distribution of particle sizes prevents 
to analyze their processes in simple ways.\cite{sizedep} 
Studies on single magnetic particles have been also performed 
but clear evidence of quantum processes has not yet been found.\cite{singlep} 

In this respect, nanoscale molecular magnets such as 
Mn$_{12}$\cite{BB1,mn1,mn2,mn3,mn4,mn5}, Fe$_8$\cite{fe1,fe2,fe3}, 
and V$_{15}$\cite{V15} 
etc. are more promising.
These molecules Mn$_{12}$ and Fe$_8$ consist of small number of atoms.
The low energy state of the system is represented by  
an effective $S=10$ spin. Because interactions 
among molecules are very small, 
each atom can be regarded as a $S=10$ single spin.
The Hamiltonian of the spin is generally given by
\begin{equation}
{\cal H}=-DS_z^2-HS_z+Q,
\label{hamS10}
\end{equation} 
where $S_z=-10,-9,\cdots 10$ and $Q$ denotes a term which causes the
quantum fluctuation, such as $S_x, S_x^2-S_y^2$, or $(S^+)^4+(S^-)^4$.
In these systems the energy levels as a function 
of the field have a discrete structure (Fig.2(a)).
There we expect an explicit quantum mechanical dynamics.
\begin{figure}
\noindent
\epsfxsize=9.0cm \epsfysize=6.5cm \epsfbox{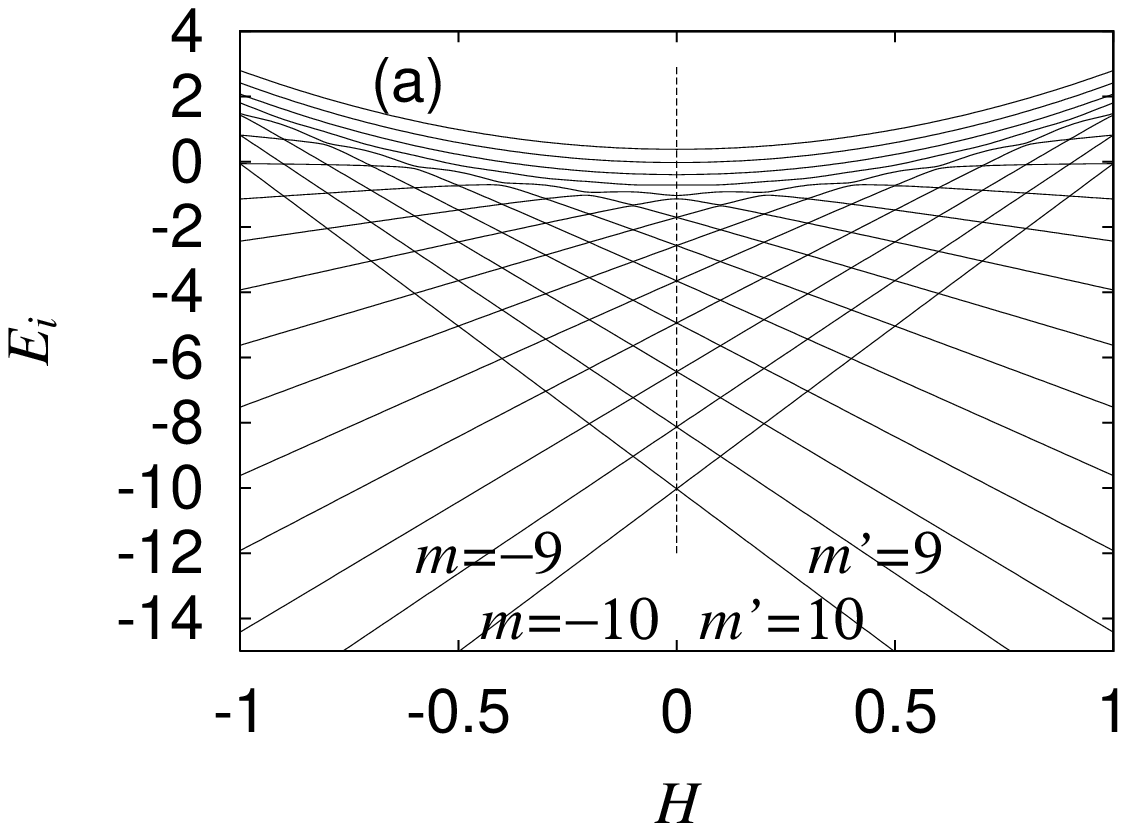} \\
\hspace*{1.2cm}
\epsfxsize=7.5cm \epsfysize=5.5cm \epsfbox{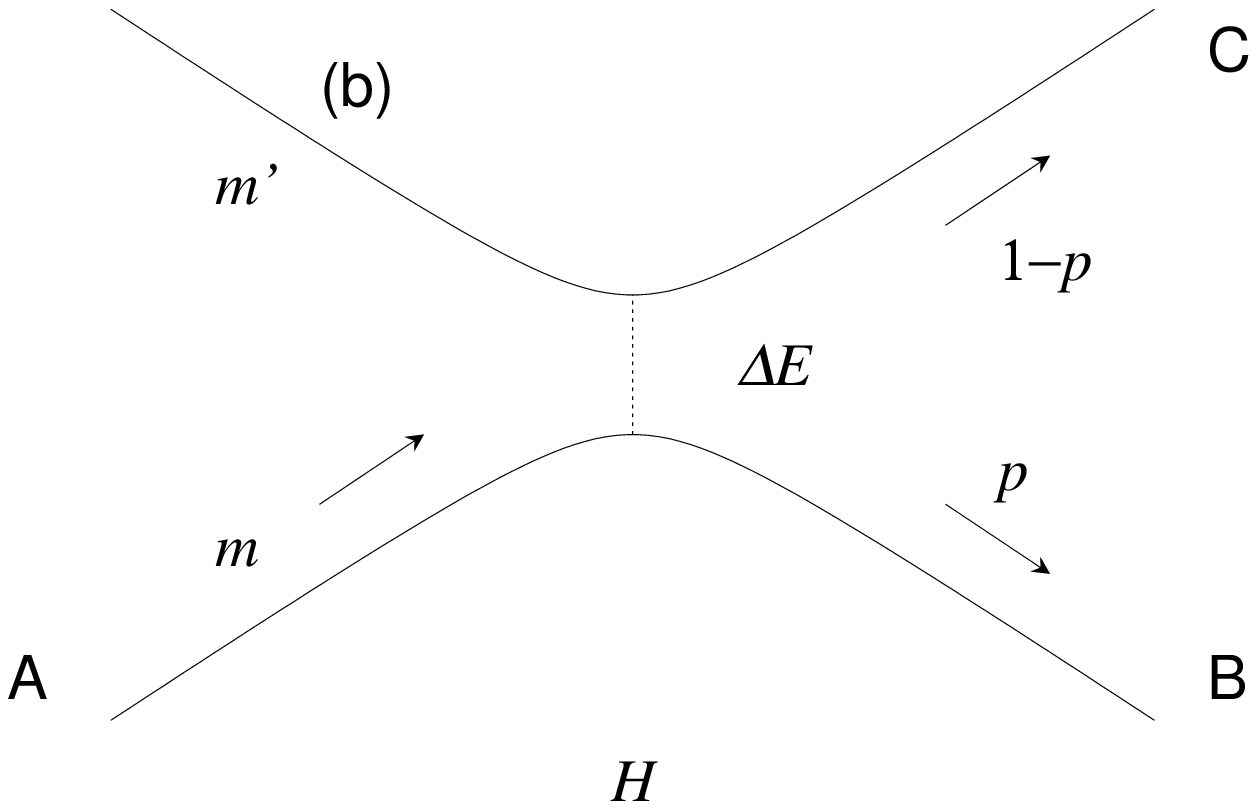}
\label{fig2h}
\caption{Energy structure of $S=10$. Uniaxial magnet as 
functions of the external field $H$. (a) global structure 
and (b) avoided level crossing.} 
\end{figure}

Due to the term $Q$, a small energy gap is formed at each crossing point 
as shown in Fig.2.
This structure is called avoided level crossing. 
When the field is swept through an avoided level crossing point, 
so called nonadiabatic transition occurs.
Nonadiabatic transition plays important roles in microscopic quantum dynamics
such as level dynamics of semiconductor, chemical reaction and optics.
Nonadiabatic transitions in various cases have been
reviewed by Nakamura.\cite{nakamura}

In the uniaxial magnets, 
the nonadiabatic transition as shown in Fig.2(b) occurs.
Here the population coming in from the channel A
is scattered to the channels B and C with probabilities  
$p$ and $1-p$, respectively. 
Here the channel B is the ground state. Thus the scattering to B is
an adiabatic change and corresponds to the tunneling.
On the other hand, $1-p$ is a probability to jump up to the channel C.
This process corresponds to the case to stay in the metastable state 
(un-tunneling).
 
The states of the channels A and C are the same state 
when there is no quantum fluctuation i.e., $Q=0$.
This unperturbed state is called the diabatic state. 
These channels A and C have similar states even in the presence of $Q$. 
The probability $p$ was studied by Landau\cite{Landau}, Zener\cite{Zener}  
and St\"uckelberg\cite{St} (LZS) and is given by 
\begin{equation}
p=1-\exp\left(-{\pi(\Delta E)^2\over 2c\hbar g\mu_{\rm B}\Delta m}\right),
\label{LZS}
\end{equation}
where $\Delta E$ is a gap at the avoided level crossing
and $\Delta m$ is the difference of magnetization of the levels,
$c$ the speed of the sweeping field $c=dH/dt$. 
Thus the term $g\mu_{\rm B}\Delta m c$ is the changing rate of 
the Zeeman energy.\cite{miya95,miya96,DMSGG} 
The probability $p$ plays an important role in 
quantum mechanical relaxation of the present system. 

In this LZS type nonadiabatic transition 
the transition occurs only in the vicinity of the crossing points. 
The first property is an essential ingredient for the relaxation at discrete
points of the magnetic field in the molecular magnets.\cite{DMSGG}   
Making use of this dependence 
we can estimate $\Delta E$ from the change of magnetization $\Delta M$  . 
Let the crossing levels have magnetizations of the corresponding adiabatic 
states $m$ and $m'$. 
The change of magnetization is given by    
\begin{equation}
\Delta M=pm'+(1-p)m-m=p(m'-m) .
\label{DeltaM}
\end{equation}
In the cases where $\Delta E$ is observed by other methods such 
as AC-susceptibility, 
this sweeping dependence of $\Delta M$ would give a method to confirm 
that $\Delta E$ really comes from the tunneling gap.\cite{miya96} 
However such confirmation has not yet been 
observed in small magnetic particles such
as magnetic dots and ferritin. 
On the other hand the two characteristics have been observed in molecular 
magnets at least qualitatively. 
As a more peculiar property of the nonadiabatic transition,
it has also been pointed out that the application of an alternate field
at the resonant points will cause a nontrivial oscillation of magnetization
due to phase interference.\cite{MSD}

In order to study experimental data  quantitatively,
we need to incorporate the effects of environments. 
Effects of noise on the LZS transition have 
been studied by Kayanuma\cite{kayanuma}.
The dynamics of molecular magnets in dissipative environments has also been
investigated quite extensively.\cite{theories}

A thermal bath causes enhancement of the relaxation, e.g. 
the thermally assisted resonant tunneling, 
where resonant tunnelings of excited states play an important role. 

In this paper we study the characteristics of the resonant tunneling
affected by thermal disturbance.
In particular, we study the effect of environment at very low temperatures 
such that relaxation process does not depend on the temperature. 
Even at such low temperatures, contact with the bath causes
relaxation between levels which have magnetizations of the same sign.
These states belong to the same valley in the potential picture of Fig. 1(a). 
The energy barrier does not exist between them.
It is found that the relaxation between them easily occurs even with
very weak disturbance.
In pure quantum mechanical motion, transitions 
between these levels are almost prohibited 
except near the avoided level crossing points.
Thus the magnetization curve with the dissipative effect  
is different from the one of a pure quantum case, although 
it does not depend on the temperature.
We call such a process "deceptive nonadiabatic transition".\cite{SMD} 

At higher temperatures, excitation levels begin to contribute 
to the relaxation phenomena. At higher temperatures, 
alternate enhancements of relaxation at resonant points are 
observed\cite{BB1}, which is called 'the parity effect'.
We consider the mechanism of such an alternation in a view point of 
nonadiabatic transition of excited states and find it as a general
property of resonant tunneling of excited states
reflecting the structure of energy levels.\cite{MSK}
We also discuss the  $\sqrt{t}$-dependence of initial decay at resonant points.
Furthermore we study various cases of the LZS process 
in fluctuating random environments.

\section{Numerical Method}

The most standard method to study quantum dynamics in dissipative 
environments is the quantum master equation (QME)
which describes the equation of motion 
of the reduced density matrix of the system $\rho(t)$
which is derived by tracing out the degrees 
of the freedom of the environment from the density matrix 
of the total system. The total system  consists of a system ${\cal H}_{\rm S}$,
a thermal bath ${\cal H}_{\rm B}$ and an interaction between them
${\cal H}_{\rm I}$:\cite{Kubo} 
\begin{equation}
{\cal H}={\cal H}_{\rm S}+{\cal H}_{\rm I}+{\cal H}_{\rm B}.
\label{hamtot}
\end{equation}
The reduced matrix is given  by
\begin{equation}
\rho(t)={\rm Tr}_{\rm B}e^{-\beta{\cal H}}.
\label{rho}
\end{equation}
We have the following equation of motion for $\rho(t)$ 
in the limit of weak coupling,
assuming that the correlation time of the bath variable 
is very short (Markovian approximation) 
\begin{equation}
{d\over dt}\rho(t)={1\over i\hbar}[{\cal H},\rho(t)]+\Gamma\rho(t),
\label{QMeq}
\end{equation}
where $\Gamma$  is a linear operator acting on $\rho(t)$. 
This equation has been used to study quantum dynamics 
of optical process, etc.
In most cases $\Gamma\rho$  has the so-called Lindblad form\cite{Lindblad}
\begin{equation}
\Gamma\rho=A^{\dagger}A\rho+\rho A^{\dagger}A 
+A^{\dagger}\rho A+A\rho A^{\dagger},
\label{Lindblad}
\end{equation}
where $A$ is an operator of the system. 
However in multileveled phenomena $\Gamma\rho$ has a more general form.
 
In the cases where the bath consists of an infinite number 
of bosons, a general expression can be derived.\cite{QME}
\begin{equation}
\frac{\partial\rho(t)}{\partial t} = 
-{\rm i} \left[{\cal H},\rho(t)\right] 
-\lambda
\left( \left[X,R\rho(t)\right] + \left[X,R\rho(t)\right]^{\dag} \right) ,
\label{CTTR}  
\end{equation}
where 
\begin{eqnarray}
\langle \bar{k} | R  | \bar{m} \rangle &=& 
\zeta (\frac{E_{\bar{k}} - E_{\bar{m}}}{\hbar})
n_{\beta} ( E_{\bar{k}} - E_{\bar{m}} )  
\langle \bar{k} | X  | \bar{m} \rangle , \nonumber \\
\zeta (\omega ) &=& I(\omega ) -I(- \omega) ,
\quad {\rm and} \quad
n_{\beta}( \omega ) =  \frac{1}{e^{\beta\omega} -1 } . \nonumber 
\end{eqnarray}
Here $\beta$ is the inverse temperature of the reservoir $1/T$,
and we set $\hbar $ to be unity. 
$| \bar{k} \rangle $ and $| \bar{m} \rangle $ are the eigenstates of 
${\cal H}$ with the eigenenergies $E_{\bar{k}}$ and $E_{\bar{m}}$, 
respectively. $I(\omega )$
is the spectral density of the boson bath.
Here we adopt the form $I (\omega ) = I_{0} \omega^{\alpha}$.
When $\alpha=1$, it corresponds to the so called Ohmic bath
and when $\alpha=2$, it corresponds to the phonon bath (super-Ohmic).
As a more realistic bath for the experimental situation at very low temperature,
we may take the dipole-field from other molecules and interactions with
the nuclear spins\cite{PS96}.
$X$ is an operator of system which is attached to bosons of the reservoir 
linearly, representing the interaction between the system and  the thermal bath. 
In the present study we take $X= \frac{1}{2} \left( S_{x} + S_{z} \right)$.
The relaxation process depends on the form of $X$. 
Generally the coupling with the transverse component $X=S_{x}$ is more efficient 
than that of the longitudinal one $X=S_{z}$ for the relaxation.
Detailed comparisons among choices of the form of the coupling
will be presented elsewhere.

For strong noise caused by fluctuating forces we can
simulate quantum dynamics 
by solving the Schr\"odinger equation in random fields.\cite{BUTSURI} 

\section{Quantum Dynamics in Dissipative Environment}
\subsection{Deceptive nonadiabatic transition}

In the lowest avoided level crossing point $(-S,S)$ 
the change of magnetization $\Delta M$ is given by (\ref{DeltaM}) . 
However at higher crossing points $(m, m')$ with $ m'<S$, 
the population scattered from $m$ to $m'$ is 
found to decay easily to the ground state, i.e.,  $m'\rightarrow  S$, 
even when the dissipative effect is so small 
that the population at the metastable level of $m$ hardly decays. 
This difference can be easily understood from the intuitive picture of Fig. 1(a). 
That is, the relaxation in the same valley, i.e., $m'\rightarrow  S$, 
is easy while the relaxation over the barrier $m\rightarrow  S$ is hard. 
In this situation, we can not apply the relaxation (\ref{DeltaM}) directly 
to estimate the LZS probability $p$. 
However we can still estimate $p$ using $\Delta  M$ 
because the relaxation from the level of $m$ occurs with 
the LZS probability 
and the relaxation to the ground state occurs in a rather short time. 
Taking these points into account, we modify the relation (\ref{DeltaM}) 
by replacing the final magnetization $m'$ by $S$:
\begin{equation}
\Delta M=pS-(1-p)m-m=p(S-m) .
\label{DeltaM2}
\end{equation}
In order to confirm these processes we performed simulations using the QME. 
First we confirmed that relaxation from the metastable point is 
unlikely to occur 
when the coupling between the bath and system is weak and the temperature 
is low.
On the other hand, a fast relaxation is observed 
between levels with magnetizations of the same sign, 
which are in the same valley in Fig. 1(a). 
Furthermore when we sweep the field we find a step-wise magnetization curve 
whose step heights do not depend on the temperature 
but are definitely different from the pure quantum case. 
In Fig. 3, we show an example of magnetization process for $T=0.1,\Gamma=0.5$
with very small effects of environment ($\lambda=0.00001$: a solid line) and 
that of pure quantum system ($\lambda=0$: dashed line).
Both of them are not temperature dependent within this temperature range.
We call this stepwise structure in dissipative environments `the deceptive
nonadiabatic transition'. 
We find that we can correctly estimate the pure quantum transition 
probabilities using the relation (\ref{DeltaM2}). 
Thus even at very low temperatures the effect of the environment 
can not be excluded, but quantum mechanical processes and dissipative effects
due to environments can be disentangled, 
and the information on the LZS probabilities can be extracted.

For the phenomena described above, 
the existence of the environment is important 
but the detailed nature is not important 
as long as it leads to fast relaxation to the ground state.
If the environment causes a change of LZS probability, 
which would be possible when the sweeping rate is very slow,
further consideration is necessary.\cite{kayanuma}
 
\begin{figure}
\noindent
\epsfxsize=9.0cm \epsfysize=6.0cm \epsfbox{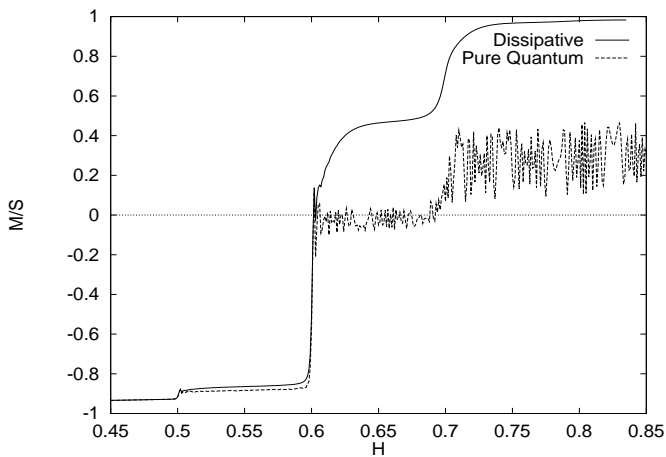} \\
\label{fig3}
\caption{Magnetization processes in pure quantum dynamics 
(a dashed line) and
a weak dissipative environment(a solid line).}
\end{figure}

\subsection{Parity effect}
At higher temperatures, excitation levels begin to contribute 
to the relaxation leading to temperature dependent phenomena. 
These processes would depend on the detailed characteristics 
of the bath and the ways of the coupling between the system and the bath.
Therefore general description is difficult. 
However here we point out a general property of relaxation
under these conditions.

As a characteristic of resonant tunneling at rather high temperatures,
it has been observed that the amount of relaxation 
at the resonant points changes alternatively.\cite{BB1}
Along a diabatic line, the energy gap increases monotonically 
as the difference of magnetizations $|m-m'|$ of levels decreases. 
Thus the transition probabilities at the resonant points increase monotonically. 
In a perturbational treatment the energy gap depends on the difference as\cite{gap}
\begin{equation}
\Delta E\propto \left({\Gamma\over D}\right)^{|m-m'|}.
\label{DeltaE}
\end{equation}
Thus we have to consider a mechanism of the alternate enhancements.
Here we interpret it from the view point of resonant tunneling of
excited state.
The transition probabilities at resonant points 
with the same value of $|m-m'|$  are nearly the same. 
Those points are located at the same horizontal level in Fig. 2. 
For example the values of $p$ given by (1.3) for the case of $\Gamma=0.45$ with 
the sweeping speed $c=0.0001$ 
at the points, 
$(-8,5), (-9,4)$, and $(-10,3)$ are 0.91, 0.64 and 0.99, respectively. 
On the other hand, those at $(-8,6), (-9,5)$ and $(-10,4)$ are 0.72, 0.037
and 0.01, which are very small. Thus most of the population at the
levels $m=-8,-9$ and $-10$ decays at the former points. These decays cause
enhancements of relaxation at $H=0.3,0.5$ and 0.7, 
which gives the parity effect.
In Fig. 4, we show the magnetization of this case with its time derivative.

Because the energy structure shown in Fig. 2 
is general for uniaxial magnets, 
we expect that the alternate enhancement of relaxation, i.e., 
the parity effect, is a general property of resonant tunneling in the thermal
environment. 
We have also pointed out that if we change the sweeping rate
the enhanced sequence is shifted. 
For example if we sweep much slower, 
the probabilities at $(-8,6), (-9,5)$ and $(-10,4)$ become large
and populations on the lines  decay there, which causes the shift
of the enhanced sequence at $H=0.2,0.4$ and 0.6.

\begin{figure}
\noindent
\epsfxsize=9.0cm \epsfysize=6.0cm \epsfbox{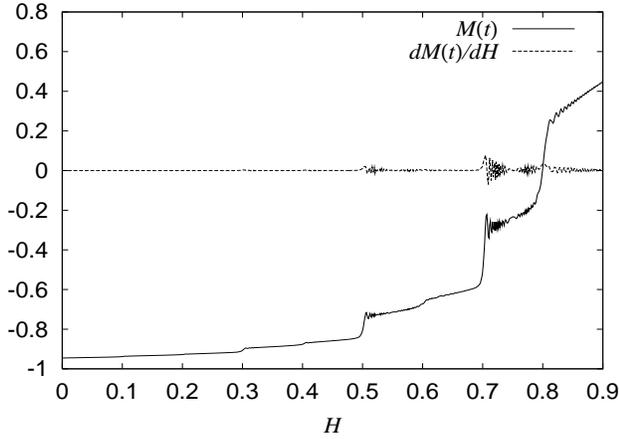} \\
\label{fig4}
\caption{The parity effect: Magnetization process (a solid line) 
and its time derivative (a dashed line).}
\end{figure}

\subsection{Non-exponential decay at the resonant point}
The magnetization which is initially polarized upward decays rather fast
at a resonant point. Here the field is set at this point and is not swept.
There are several paths for the magnetization to relax at this point. 
First let us consider relaxation by the nonadiabatic transitions 
at the lowest resonant point. 
Because at the resonant point the energy gap is very narrow, 
the field fluctuates around the point as shown in Fig. 5(a). 
If we regard the motion of the field as a Brownian motion, 
it is known that the number of the times 
the field crosses the resonant point is proportional to $\sqrt{t}$,
i.e., the recurrence time of one-dimensional Brownian motion.\cite{BM}. 

\begin{figure}
\noindent
\epsfxsize=4.5cm \epsfysize=5.3cm \epsfbox{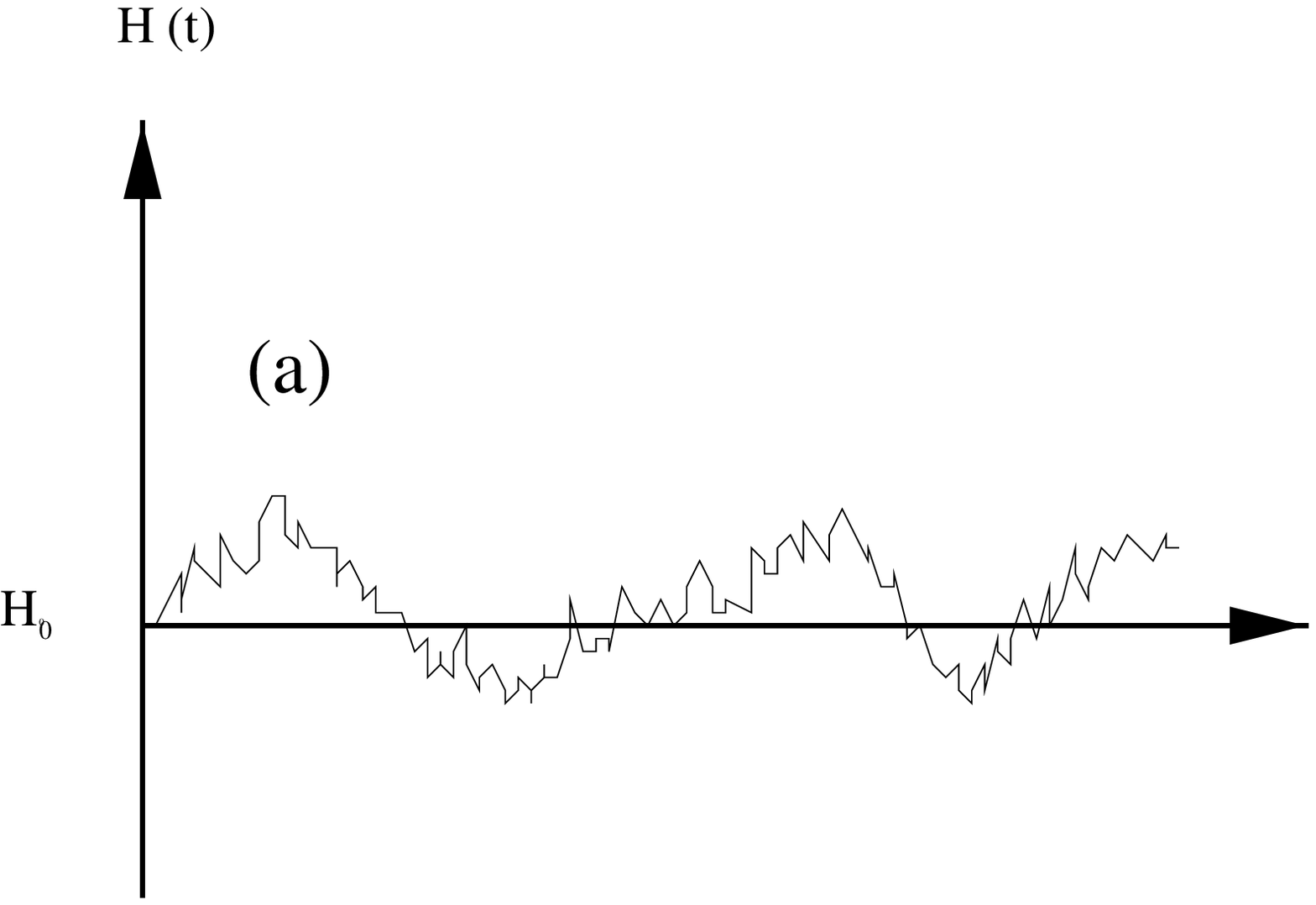}
\epsfxsize=4.5cm \epsfysize=5.3cm \epsfbox{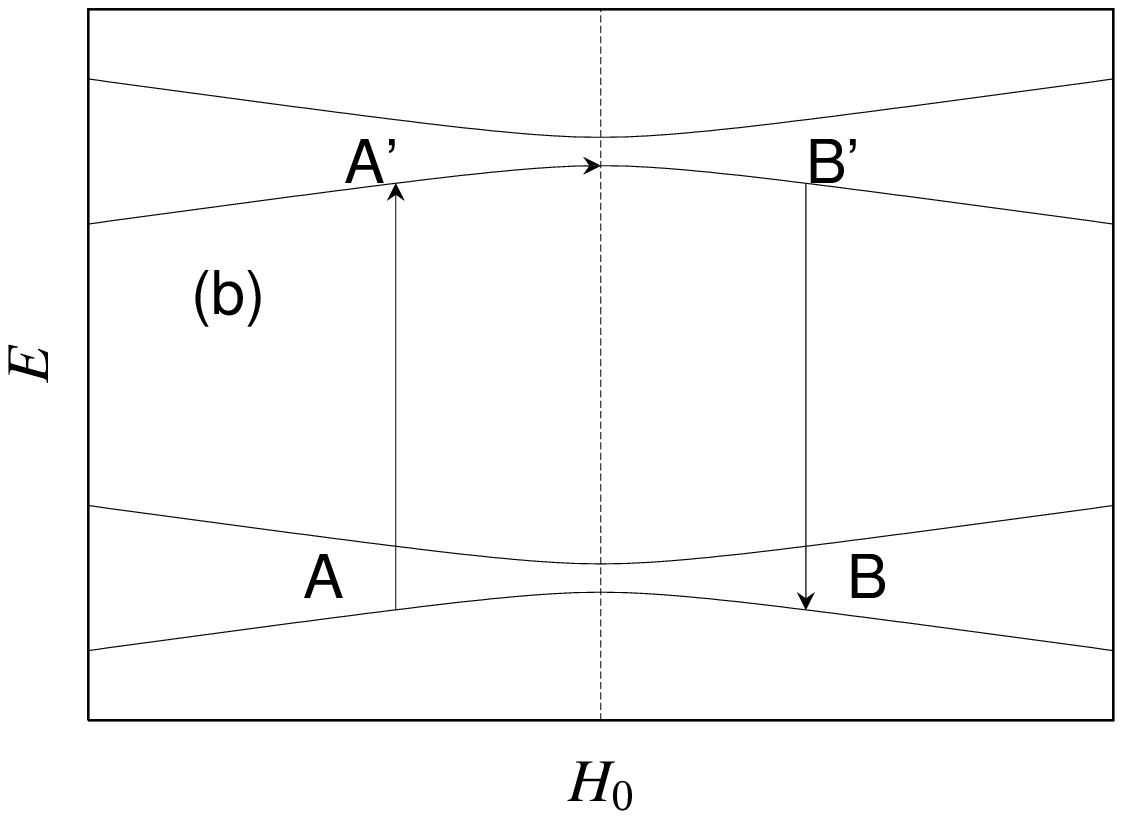} 
\caption{(a) Random field near the resonant point and (b) Relaxation process
through the excited state.}
\end{figure}

At each crossing, the population moves to the other branch 
by the LZS transition probability
\begin{equation}
p_i=1-\exp(-{\pi(\Delta E)^2\over 4c_i})\simeq{\pi(\Delta E)^2\over 4c_i}.
\label{pi}
\end{equation}
In a pure quantum mechanical process, 
quantum mechanical interference occurs among the transitions.\cite{MSD} 
But assuming a fast decoherence,
the total transition probability is expected to be given by
\begin{equation}
p_{\rm total}(t)\simeq\sum_ip_i\simeq \sqrt{t}\left\langle{1\over c_i}\right\rangle
{\pi(\Delta E)^2 \over 4}\equiv \alpha \sqrt{t}.
\label{ptotal}
\end{equation}
Thus we naturally expect that the magnetization decays as 
\begin{equation}
\Delta M= M_0(1-2\alpha\sqrt{t})
\label{Mint}
\end{equation}
at the initial stage.

For a longer time scale, the field does not fluctuate freely 
but is confined near the resonant point. 
Thus $p_{\rm total}$ for long time is proportional to $t$.
Therefore the magnetization decays in exponentially
\begin{equation}
M(t)\simeq e^{-t/\tau},\quad \tau={p_{\rm total}(t)\over t}.
\label{Mt}
\end{equation}

This mechanism described above may give the simplest explanation 
of the $\sqrt{t}$ behavior. 
A more detailed analysis has been given in the reference \cite{PS96}, 
taking into account the explicit nature of the fluctuation of the external field.

\subsection{Modification of the transition rate}

When the amplitude of the external disturbance is strong, 
we have to use another estimation of the transition probability.
Kayamura and Nakayama have investigated LZS transition in fluctuating field 
and obtained expression of the transition probability. \cite{kayanuma} 
In the case where the sweeping rate is slow and the transient time  through
the resonant point is much longer than the phase coherence time, then they obtained 
the transition probability as
\begin{equation}
p=p_{\rm SD}={1\over2}(1-\exp(-\pi(\Delta E)^2/2c\hbar g\mu_{\rm B}\Delta m)).
\label{Kayanumaeq}
\end{equation}
On the other hand, when the transient time is very short, transition
probability does not change from that of the pure LZS transition
\begin{equation}
p=p_{\rm LZS}=1-\exp(-\pi(\Delta E)^2/4c\hbar g\mu_{\rm B}\Delta m).
\label{Kayanumaeq2}
\end{equation}
They confirmed such dependences by numerical simulation.

Next, let us consider a case where excited levels contribute to 
the relaxation. 
If the frequency of contacts between the system and the bath is high, 
a tunneling through the excited state would enhance 
the relaxation rate even the population at the excited state is very little.
\cite{WWnoise}

Let us consider the case where
the LZS transition probability at the lowest level $p_0$ 
is very small and the one at the excited level $p_1$ is of the order 1. 
Let us consider a case where a state is excited to the exited level. 
In the off-resonant region this excited state decays to the original state 
very rapidly. The population at the excited state $n_{\rm E}$ 
is determined by the balance equation
\begin{equation}
n_{\rm E}R_{{\rm E}\rightarrow {\rm G}}=n_GR_{{\rm G}\rightarrow {\rm E}},
\label{balance}
\end{equation}
where $n_{\rm G}$ is the population in the ground state 
and $R_{{\rm E}\rightarrow {\rm G}}$ and 
$R_{{\rm G}\rightarrow {\rm E}}$ are 
transition rate from the excited state to the ground state 
and vise versa. 
Although
$n_{\rm E}=R_{{\rm G}\rightarrow {\rm E}}/R_{{\rm E}\rightarrow {\rm G}}n_G$
is very small,  $R_{{\rm E}\rightarrow {\rm G}}$ and 
$R_{{\rm G}\rightarrow {\rm E}}$ themselves can be very large.

At the resonant point, 
small fluctuations of the field would cause the crossing point
as we see in Fig. 5(a). 
Thus the population pumped to the excited state A'
can be transferred to the state B' 
and then it decays to B instead of A. (Fig.5(b))  
This path (A$\rightarrow$ A'$\rightarrow$ B'$\rightarrow$ B)  
becomes dominant when $p_1\gg p_0$ 
and fluctuation of the field is rather fast. 
The opposite path (B$\rightarrow$ B'$\rightarrow$ A'$\rightarrow$ A) is also larger. 
In the initial stage most of the population 
is at A and therefore the population moves along the former path. Thus  
the effective transition rate A $\rightarrow$ B is enhanced very much. 
\begin{equation}
p_{\rm eff}\simeq R_{{\rm G}\rightarrow {\rm E}}\nu p_1,
\label{peff}
\end{equation}
where $\nu$ is the frequency of the crossing. 
For a short time $\nu$ is proportional to $\sqrt{t}$ and for a long time 
it is proportional to $t$ as we saw above.

To study the transition probability in this case, 
the correct information for $R_{{\rm G}\rightarrow {\rm E}}$ and $\nu$ 
is necessary, but it is generally difficult.
However, if we could estimate these quantities from
the enhancement, it would yield detailed knowledge of the bath.
In nanoscale molecular magnets, 
it would be possible to study such a detailed property, 
which is a very interesting research area in the future.

\section{Summary and Discussion}

Nanoscale molecular magnets display several phenomena 
which originate from explicit quantum mechanical transitions 
between discrete levels. 
In this paper we studied effects of dissipative environment 
which smears out the pure quantum processes.
So far the relaxation processes have been a kind of black-box 
and have been treated only phenomenologically.  
But it would be possible to begin to study explicit processes of relation 
in nanoscale magnets because of their simple form. 

So far we studied the $S=10$ spin representing the low energy structure
of magnetic energy levels.
Let us consider the structure of the full energy level.
The molecular magnets have complicated structures. For example Mn$_{12}$
includes 12 Mn molecules with many other atoms which have nuclear spins. 
Thus the dimension of total Hamiltonian is $5^8\cdot 4^4\times I$, 
where $I$ comes from the degree of freedom of nuclear spins.
This degree of  nuclear spins causes random effects on each Mn atom.
It would be an interesting problem to study how this random field on
individual atoms causes changes of the energy levels at low temperatures.

Even without the effects of nuclear spins, 
there are dipole-dipole couplings among the molecules 
which cause random noise on the whole molecule.
Thus it should be taken into account even in the view point of $S=10$
spin. Effects of this field are studied as `feedback-effect' on the 
LZS process of magnetization.\cite{HDMS}  
It would be interesting to study natures of noises explicitly in
nanoscale molecular magnets.

\section*{Acknowledgements}
\vspace*{-0.1cm} 

 We would like to thank Professors Bernard Barbara, W. Wernsdorfer,
Y. Kayanuma, H. Nakamura and H. Shibata for encouraging discussions.
We also thank the Grant-in-Aid from the Ministry of Education,
Science and Culture for the international collaboration program, which
has helped the work discussed in the present article. 
The simulations have been made using the computational facility of the
Super Computer
Center of Institute for Solid State Physics, University of Tokyo, which 
is also appreciated. 


\vspace*{-0.1cm} 

\begin{thebibliography}{99}
\bibitem{L6769} J. S. Langer: Ann. Phys. {\bf 41} (1967) 108, 
{\bf 54} (1969) 258.
\bibitem{Fisher}  M. E. Fisher: Physics, {\bf 3} (1967) 255 and 
in {\it Proceeding of the Gibbs Symposium, Yale University},
ed. G. D. Mostow and D. G. Caldi: (American Mathematical Society, Providence, 
RI, 1990).
\bibitem{RTMS94}
P. A. Rikvold, H. Tomita, S. Miyashita, and S. W. Sides: Phys. Rev. E 
{\bf 49} (1994) 5080.
\bibitem{coherent} L. N\'eel: Ann. Geophys. {\bf 5} (1949) 99.
W. F. Brown: Phys. Rev. {\bf 130} (1963) 1677.
L. Gunther and B. Barbara: Phys. Rev. B {\bf 49} (1994) 3926.
\bibitem{ferritin} D. D. Awschalom, M. A. MaCord and G. Grinstein:
Phys. Rev. Lett. {\bf 65} (1990) 783, {\bf 68} (1992) 3092 and {\bf 71} (1993) 4279.
S. Gider et al.: J. Appl. Phys. {\bf 79} (1996) 5324.
J. Tejada, et al.: Phys. Rev. Lett. {\bf 79} (1997) 1754.
\bibitem{sizedep}
B. Barbara and W. Wernsdorfer: Current Opinion in Solid State, {\bf 2} 
(1997) 220.
\bibitem{singlep}
W. Wernsdorfer, et al.: Phys. Rev. Lett. {\bf 78} (1997) 1791.
W. Wernsdorfer, et al.: Phys. Rev. Lett. {\bf 79} (1997) 4014.
M. C. Miguel and E. M. Chudnovsky: Phys. Rev. B {\bf 54} (1996) 388.
\bibitem{BB1} B. Barbara, L. Thomas, F. Lionti, I. Chiorescu and A. Sulpice:
J. Mag. Mag. Mat. {\bf 200} (1999) 167.
\bibitem{mn1} J. R. Friedman and M. P. Sarachik, T. Tejada and
R. Ziolo: Phys. Rev. Lett. {\bf 76} (1996) 3830.
\bibitem{mn2} L. Thomas, F. Lionti, R. Ballou, D. Gatteschi,
R. Sessoli and B. Barbara: Nature {\bf 383} (1996) 145.
\bibitem{mn3} L. Thomas et al.: Nature {\bf 383} (1996) 145.
\bibitem{mn4} F. Lionti, L. Thomas, R. Ballou, Barbara, A. Sulpice,
R. Sessoli and Gatteschi: J. Appl. Phys. {\bf 81} (1997) 4608.
\bibitem{mn5} J. M. Hernandez, X. X. Zhang, F. Luis, and T. Tejada,
J. R. Friedman, M. P. Sarachik and R. Ziolo: Phys. Rev. B {\bf 55} 
(1997) 5858. J. A. A. J. Perenboom, J. S. Brooks, S. Hill, T. Hathaway, and
N. S. Dalal, Phys. Rev. B {\bf 58} (1998) 330.
\bibitem{fe1} C. Sangregorio, T. Ohm, C. Paulsen, R. Sessoli, D. Gatteschi: 
Phys. Rev. Lett. {\bf 78} (1997) 4645.
\bibitem{fe2} W. Wernsdorfer and R. Sessoli: Science {\bf 284} (1999) 133.
\bibitem{fe3} W. Wernsdorfer et al.: Phys. Rev. Lett. {\bf 82} (1999) 3909.
\bibitem{V15} D. Gattesschi, et al.: Letter to Nature {\bf 354} (1991) 12.
A. Barra, et al.: J. Am. Chem. Soc. {\bf 114} (1992) 8509.
D.  Gattesschi, L. Pardi, A. Barra and A. Muller: Molecular Engineering, 
{\bf 3} (1993) 157.
\bibitem{nakamura} H. Nakamura: {\em Dynamics of Molecules and 
Chemical Reactions} (Marcel Dekker, 1996) 473.
\bibitem{Landau} L. Landau: Phys. Z. Sowjetunion {\bf 2} (1932) 46.
\bibitem{Zener} C. Zener: Proc. R. Soc. London Ser. A {\bf 137} (1932) 696.
\bibitem{St} E. C. G. St\"uckelberg: Helv. Phys. Acta {\bf 5} (1932) 369.
N. S. Dalal: Phys. Rev. B (1998) 330.
\bibitem{miya95} S. Miyashita: J. Phys. Soc. Jpn. {\bf 64} (1995) 3207.
\bibitem{miya96} S. Miyashita: J. Phys. Soc. Jpn. {\bf 65} (1996) 2734.
\bibitem{DMSGG}
H. De Raedt, S. Miyashita, K. Saito, D. Garc\'{i}a-Pablos, and N. Garc\'{i}a:
Phys. Rev. B, {\bf 56} (1997) 11761.
 \bibitem{MSD}
 S. Miyashita, K. Saito, and H. De Raedt: Phys. Rev. Lett. {\bf 80} 
(1998) 1525. 
\bibitem{kayanuma} Y. Kayanuma and H. Nakamura: Phys. Rev. B {\bf 57} (1998) 13099.
\bibitem{theories} D. A. Garanin and E. M. Chudovsky: Phys. Rev. B {\bf 56} (1997) 11102. A. Fort, A. Rettori, J. Villain, D. Gatteschi, and R. Sessoli: Phys. Rev. Lett. {\bf 80} (1998) 612. F. Luis, J. Bartolom\'{e}, and F. Fern\'{a}ndez: Phys. Rev. B {\bf 57} (1998) 505. M. N. Leuenberger and D. Loss: cond-mat/9911065/.
\bibitem{SMD}
K. Saito, Miyashita, and H. De Raedt: Phys. Rev B {\bf 60} (1999) 14553.
\bibitem{MSK} S. Miyashita, K. Saito and H. Kobayashi: cond-mat/9911148. 
\bibitem{Kubo} R. Kubo, M. Toda, and N. Hashitsume: 
{\em Statistical Physics II} (Springer-Verlag, New York, 1985). 
W. H. Louisell: {\em Quantum Statistical Properties of Radiation} 
(Wiley, New York, 1973). W. Weidlich and F. Haake: Z. 
Phys.\  {\bf 185}, (1965) 30.
\bibitem{Lindblad} G. Lindblad: Commun. Math. Phys. {\bf 48} (1976) 119.
\bibitem{QME}
K. Saito, S. Takesue, S. Miyashita:  to be published in Phys. Rev. E, 
cond-mat/9810069.
\bibitem{PS96} N. V. Prokof'ev and P. C. E. Stamp: 
Phys. Rev. Lett. {\bf 80} (1998) 5794 and J. Low Temp. Phys. {\bf 113} (1998) 1147.
\bibitem{BUTSURI} S. Miyashita: BUTSURI {\bf 53} (1998) 259 (in Japanese).
\bibitem{gap} D. A. Garanin: J. Phys. A {\bf 24} (1991) L61.
\bibitem{BM} W. Feller: {\em An Introduction to Probability Theory and Its Applications} (1957, 
John Wiley \& Sons, New York). 
\bibitem{WWnoise} W. Wernsdorfer: private comunication.
\bibitem{HDMS} A. Hams, H. De Raedt, S. Miyashita and K. Saito: unpublished.
\end{thebibliography}
\end{document}